\documentclass[aps,prl,reprint]{revtex4-1}

\usepackage{graphicx}
\usepackage{epsfig}
\usepackage{dcolumn}
\usepackage{bm}
\usepackage{color}
\usepackage{amsmath}
\usepackage{amsfonts}
\usepackage{amssymb}
\usepackage{epstopdf}
\usepackage{graphics}
\usepackage{multirow}
\usepackage{multirow,bigdelim}
\newcommand{\tick}{$\checkmark$}
\newcommand{\cross}{$\mathbf{\times}$}

\def\be{\begin{equation}}       \def\ee{\end{equation}}
\def\bea{\begin{eqnarray}}      \def\eea{\end{eqnarray}}
\def\ba{\begin{array} }
\def\ea{\end{array} }
\def\bnum{\begin{enumerate} }
\def\enum{\end{enumerate}}

\def\=>{\Rightarrow}
\def\>{\rightarrow}

\def\eye2{Fathbb{I}}

\renewcommand{\>}{\rangle}

\newcommand{\vdW}{\text{vdW}}

\usepackage[normalem]{ulem}
\usepackage{color}
\usepackage{xcolor}

\definecolor{Mygrey}{gray}{0.80}

\ifdefined\NoComments

\else
 \ifdefined\LightComments

 \else

 \fi
\fi

\begin{document}
\title{Evaluation of van der Waals density-functionals for layered materials}

\author{Sherif Abdulkader Tawfik}
\affiliation{School of Mathematical and Physical Sciences, University of Technology Sydney, Ultimo, New South Wales 2007, Australia}
\email{sherif.abbas@uts.edu.au}
\author{Tim Gould}
\affiliation{
	Qld Micro- and Nanotechnology Centre, Griffith University, Nathan, Qld 4111, Australia
}
\email{t.gould@griffith.edu.au}
\author{Catherine Stampfl}
\affiliation{School of Physics, The University of Sydney, New South Wales, 2006, Australia}
\author{Michael J. Ford}
\email{mike.ford@uts.edu.au}
\affiliation{School of Mathematical and Physical Sciences, University of Technology Sydney, Ultimo, New South Wales 2007, Australia}

\begin{abstract}
  In 2012, Bj\"orkman \emph{et al.} posed the question ``Are we van der
  Waals ready?'' [J. Phys.: Condens. Matter, 2012, 24, 424218]
  about the ability of \emph{ab initio} modelling to reproduce van der
  Waals (vdW) dispersion forces in layered materials. The answer at that
  time was no, however.
  Here we report on a new generation of vdW dispersion
  models and show that one, fractionally-ionic atom (FIA) theory
  with many-body dispersions,
  offers close to quantitative predictions for layered structures.
  Furthermore, it does so from a qualitatively correct picture of
  dispersion forces. Other methods, such as D3 and optB88vdW also
  work well, albeit with some exceptions. We thus argue that we are nearly
  vdW ready, and that some modern dispersion methods are
  accurate enough to be used for nanomaterial prediction, albeit
  with some caution required.
\end{abstract}

\maketitle

\section{Introduction}
van der Waals (vdW) heterostructures \cite{Geim2013}, and nanoscience
more generally, promise to transform science and technology by
offering controllable material properties at the nanoscale. But many
challenges must be met for the unprecedented benefits of
heterostructures to be realised in technology. Not least of these is
understanding what combinations of two-dimensional (2D) layers are
both useful and structurally stable and, relatedly, how we can
engineer structures to improve stability.

Significant work must thus be carried out to devise useful
heterostructures, especially working out what combinations of layer
types are useful and feasible. Isolating 2D layers is difficult,
however. Assembling heterostructures is more
difficult \cite{Geim2013}. Thus, studying even a small representative
space of interesting 
heterostructure seems like an impossible task for experimental
laboratories.

Conveniently, heterostructure science has been paralleled by advances
in computer modelling \cite{Gould2016-Chapter,Paul2017-2DComp},
which offers the ability to scan large spaces
of candidate materials quickly and efficiently.
Prediction of heterostructure properties relies, at a minimum,
on two major factors: an ability to reproduce lattice parameters, and
thus basic geometries; and an ability to reproduce energies and their
differences, and thus to understand the relative stability of
different geometries. A good method must thus be able to reproduce
these properties if it is to offer reliable results. Otherwise time
can be wasted by experimentally exploring poor candidates misidentified
as good by the virtual screening process. More worryingly, good
candidates might never make it past the virtual screening process.
Both hamper technological progress.

In this Letter we report an assessment of modern vdW
dispersion approaches on a representative sample of 2D materials,
including graphene, boron nitride, MoS$_2$, MoSe$_2$, MoTe$_2$,
WS$_2$, PdTe$_2$, TaS$_2$, TaSe$_2$, HfS$_2$, HfSe$_2$, HfTe$_2$. Our
tests, building on previous work by Bj\"orkman and
coworkers\cite{Bjorkman2012,Bjorkman2012-Review}
are designed to interrogate how well modern approaches can deal
with the most basic properties of heterostructures.
They thus include an important additional test
not previously considered by Bj\"orkman \emph{at al}: the quality of
energy differences between different structural arrangements of
homostructures. This test is critically important as it shows that a
method not only works well in optimal homostructures, where it
may benefit from a cancelattion of errors at the opimal interlayer
spacing, but is also likely to
work well in heterostructures which, due to the presence of
incommensurate lattices, involve layered structures and their
interactions in a range of relative
configurations\cite{Leconte2017-Moire}.

We compare the predictive accuracy of 11 modern vdW 
methods against the predictions of the random-phase
approximation (RPA) \cite{Eshuis2012,Dobson2012-JPCM,Ren2012} which
have been established as one of the most accurate methods for
describing the physics of vdW materials
\cite{Dobson2014-IJQC,AdhesionSachs}. \textbf{Quantum Monte-Carlo
  methods (QMC), widely considered to be highly accurate,
  have been applied for predicting the interlayer distance and the
  binding energy of graphite \cite{QMC1,QMC2}. QMC, however,
  has only been applied to a limited number of systems, while RPA
  has been applied for a wide range of systems.} Our tests show that
the fractional ionic atoms method \cite{Gould2016-FI} (referred to
here as FIA for notational brevity -- the method is more fully
described as MBD@rsSCS/FI+ER as per the original paper) achieves a
useful balance between the accurate prediction of the lattice
constants, energies, and energy differences.

\begin{figure}[t]
  \includegraphics[width=\linewidth]{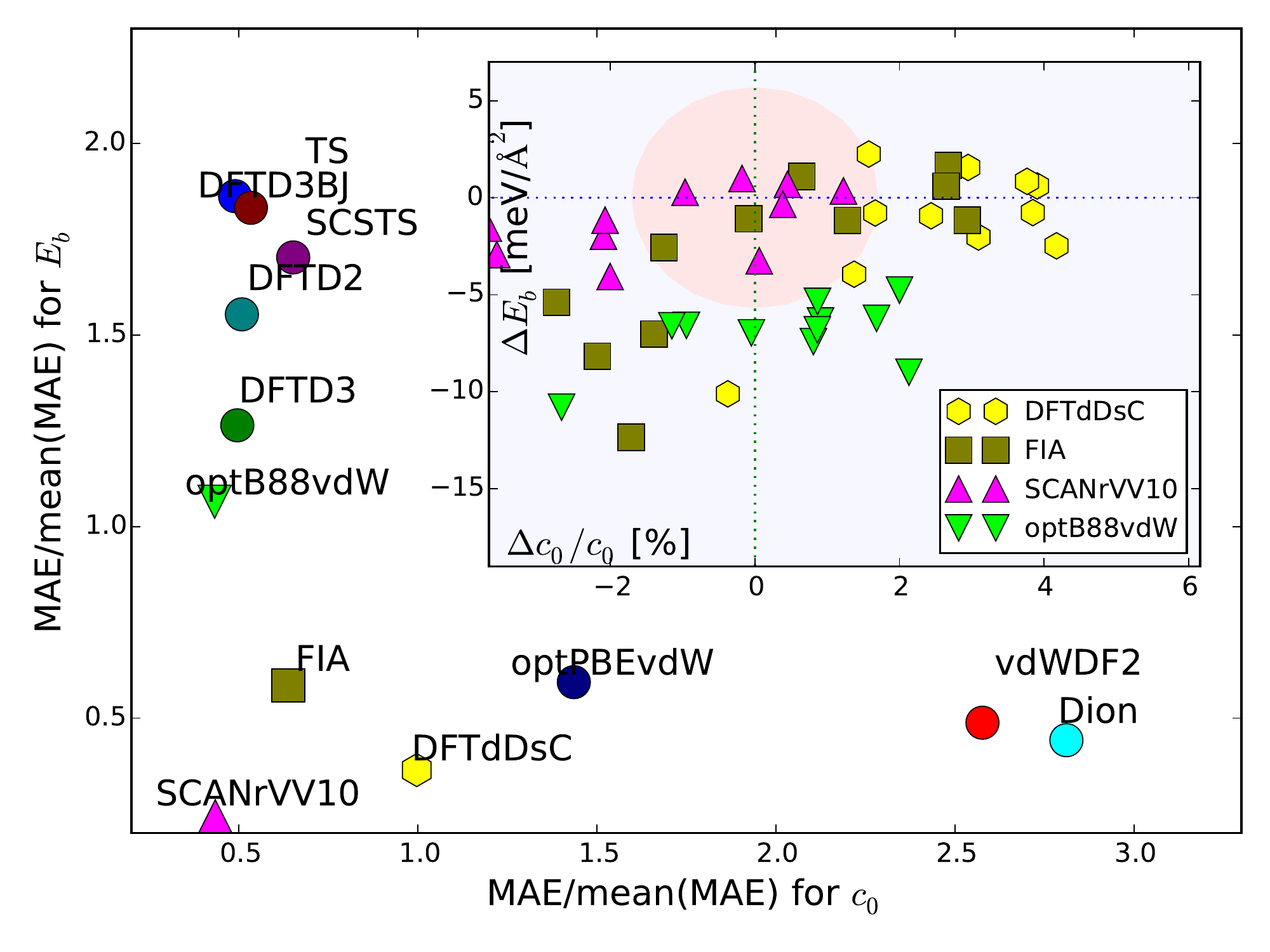}
  \caption{(Main) Plot of the normalised mean average error (MAE) in
    the binding energy and lattice spacing for all tested methods.
    (Inset) Scatter plot of the four best methods showing all
    tested materials. The oval indicates the optimal goal of
    $\pm 2\%$ for $c_0$ and $\pm 5$~meV/fu for $E_b$.
    \label{fig:Rankings}}
\end{figure}

\begin{figure}[t]
  \includegraphics[width=\linewidth]{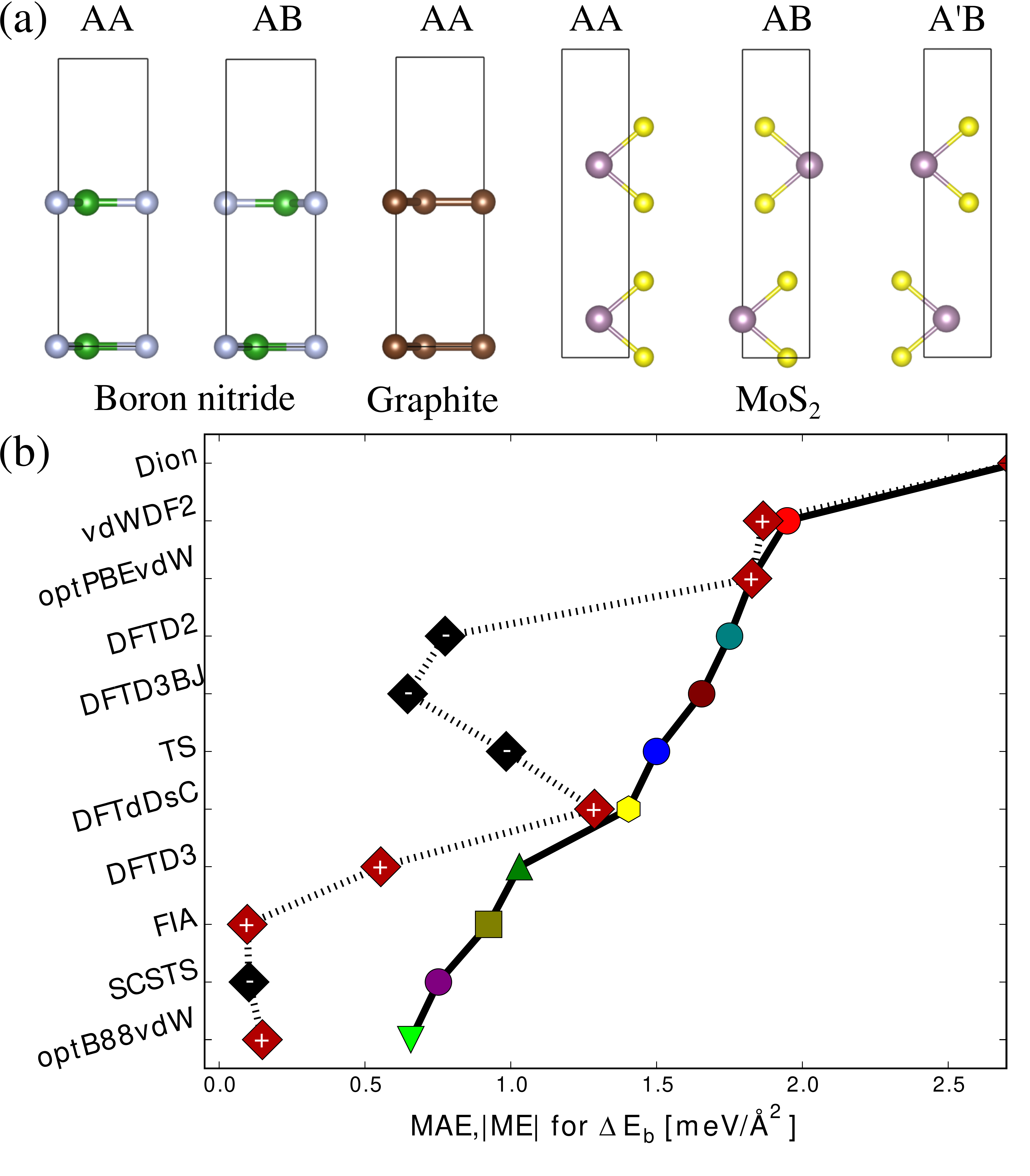}
  \caption{(a) The side-view of the various stacking orders for
    graphene, hBN and MoS$_2$. (b) Plot of the MAE (thick line and
    markers) and ME (dashed line, with indicators of sign) for
    energetic differences $\Delta E=E_0^{G1}-E_0^{G2}$ that serve as a
    proxy quality metric for heterostructures.
    Here $G1$ and $G2$ indicate different stackings of the same material.
    \label{fig:Stacking}
  }
\end{figure}

\section{Theory}
Dispersion forces, or van der Waals, forces, are weak forces that
arise from the coupling between charge fluctuations in quantum
systems. There has been a steady improvement in \emph{ab initio}
methods which 
account for a full description of chemical and dispersive forces.
Broadly speaking, these fall into three categories: semi-empirical
(SE) ``DX'' models from Grimme~\emph{et al.} \cite{Grimme-D2,Grimme-D3,D3BJ},
models based on atomic polarizabilities (AP-D) modified by
the electron density \cite{Tkatchenko-Scheffler,MBD,
  dDsC,Gould2016-FI},
and full density functional approximations (DFA)
based on pairwise dispersion models
using only the density \cite{klimevs2009chemical,Dion}.
Table~\ref{tab:methods} summarises the methods
applied in this Letter, categorized according to the above scheme.
Further details of all methods are provided in the original works, and
more detailed summaries of modern dispersion approaches are provided
in Refs.~\cite{Gould2016-Chapter,Hermann2017-Review}.
Here we focus on methods which correct generalized gradient
approximations, as these are the most widely available and easily
employed class of density functionals.

To understand the qualitative advantages and disadvantages of
dispersion methods, we must first focus on the competing pictures of 
dispersion forces: the first, more dominant amongst chemists, is that
interactions 
between atoms can be modified and then summed to get the total
interaction. The second, more dominant amongst physicists, employs
polarizability models from Lifshitz which are based around the physics
of macroscopic solids.

Dobson recently described how these two pictures can be connected to
one another \cite{Dobson2014-IJQC} and thus applied to nanostructures
which share properties with molecules and bulk solids.
He divided contributions to dispersion forces into 
three types of ``non-additivity'' effects. His first, here called
``Dobson-A'' effects, involves contributions from chemical
environments and is present in all useful theories of dispersion
forces. His second, Dobson-B effects, involves many-body
interactions known from Lifshitz theory and is present in the
RPA, and the Many-body dispersion (MBD) \cite{MBD} class of
approximations which has been used to show their vital
importance for describing nanostructure
binding \cite{DiStasio2012,Gobre2013,Ambrosetti2016}.
His third, Dobson-C, involves metallic/insulating physics and
is approximated in MBD \cite{DiStasio2012} and fully present in RPA.
Dobson-C effects mostly affect asymptotic
physics \cite{Dobson2006,Lebegue2010,Dobson2014-1,Dobson2016-LRT}
[that is, the energetic behavior of layered materials for large
($\gg10$~nm) separation distances between the layers]
and are unlikely to be relevant to typical studies of
two-dimensional heterostructures.

\begin{table}
  \caption{\label{tab:methods} The computational methods used in this
    study, the key reference of each method, the classification of the
    method [whether based on the semi-empirical methods of Grimme (S-E),
    atomic polarizabilities modifed by the density (AP-D)
    or pure functionals of the electronic density (DFA), which
    Doboson nonadditivity types\cite{Dobson2014-IJQC} are
    supported [? means partially supported].
    The final columns ``$c_0$'' and ``$E_b$ group the methods
    by their success in predicting $c_0$ and/or $E_b$. The FIA
    method is the one that is closest to the accuracy of the RPA
    method.}
    \begin{ruledtabular}\begin{tabular}{cc|c|ccc|cc}
        Method  &  Ref.  & Class & \multicolumn{3}{c|}{Non-additivity} &
        \multicolumn{2}{c}{Quality} \\
        & & & A? & B? & C? & $c_0$ &$E_b$ \\
        \hline
        RPA & \cite{RPA-Harl} & ACFD & \tick & \tick & \tick &
\rdelim\}{3}{30pt}[\tick] & \rdelim\}{3}{30pt}[\tick] \\
        SCAN-rVV10 & \cite{SCANrVV10} & DFA & \tick & \cross & \cross & & \\  
        FIA & \cite{Gould2016-FI} & AP-D & \tick & \tick & ? & & \\  
        \hline
        TS & \cite{Tkatchenko-Scheffler} & AP-D & \tick & \cross & \cross& \rdelim\}{6}{30pt}[\tick]&\rdelim\}{6}{30pt}[\cross] \\
        DFTD3BJ& \cite{D3BJ} & S-E & \tick & ? & \cross& & \\  
        SCSTS & \cite{MBD} & AP-D & \tick & \tick & \cross & & \\ 
        DFTD2 & \cite{Grimme-D2} & S-E & \tick & \cross & \cross & & \\
        DFTD3 & \cite{Grimme-D3} & S-E & \tick & ? & \cross & &\\
        optB88vdW & \cite{klimevs2009chemical} & DFA & \tick & \cross & \cross & &\\
        \hline
        DFTdDsC & \cite{dDsC} & AP-D & \tick & \cross & \cross & \rdelim\}{4}{30pt}[\cross] &\rdelim\}{4}{30pt}[\tick]  \\
        optPBEvdW & \cite{klimevs2009chemical} & DFA & \tick & \cross & \cross & & \\
        vdWDF2 & \cite{vdWDF2} & DFA & \tick & \cross & \cross& & \\
        Dion & \cite{Dion} & DFA & \tick & \cross & \cross & &\\
    \end{tabular}\end{ruledtabular}
\end{table}

In the computational assessment of the quality of the vdW methods, a
critical concern is the identification of benchmarks. In such studies,
the two primary quantities that are predicted are the $c_0$ lattice
parameter and the layer binding energy, $E_b$, and these must be
tested against suitable benchmark data. Lattice parameters are known
accurately from experiments, which can serve as a benchmark, at least
up to contributions from the zero-point energy. The situation for
binding energies $E_b$ is rather more challenging, however, given the
inaccuracies encountered in the indirect measurement of small energy
differences. Therefore, for $E_b$, we use instead published RPA
values for
$E_b$ \cite{Lebegue2010,Bjorkman2012,He2014,Leconte2017-Moire}
as benchmarks for the present study, as done in previous
studies \cite{Bjorkman2012}. Note that the RPA gives good agreement
with experiment for lattice constants\cite{Harl2009,Bjorkman2012} and
includes all of Dobson's nonadditivity classes A, B and C.
It is thus likely to carry a complete picture of binding in layered
materials.




With the qualitative picture, and quantitative benchmarks
established, we can now consider the models in our study.
All three categories (SE, AP-D and DFA) of dispersion methods have
seen steady improvements in accuracy over the past decade
\cite{Bjorkman2012-Review}. Table~\ref{tab:methods} reports the
list of approaches tested here, representing recent iterations
in each category (we note that MBD@rsSCS has problems in transition
metal dichalcogenides\cite{Gould2016-FI} and was thus excluded from
our studies).

Any new method is assessed by performing statistical tests
on the outcomes of calculations for a set of benchmark systems,
compared against higher-level theory (or experimental) data.
\textbf{Usually, such benchmarking has lacked the inclusion of 2D heterostructures, and has been based only
on atomic and molecular systems.} Tests are typically reported only for
graphite and hexagonal boron nitride, if anything.
Methods may thus suffer from inaccuracies in predicting the
properties of the interaction in 2D materials generally.
Here, we seek to remedy this deficiency to understand which methods
are suitable for calculations of 2D and related systems.

At the pairwise level, we test Grimme's D2 empirical
correction \cite{Grimme-D2} (DFTD2), Grimme's D3 empirical
correction \cite{Grimme-D3} in its original form (DFTD3) and
with Becke-Johnson damping (DFTD3BJ), the exchange-hole based
correction of Steinmann and Corminboeuf \cite{dDsC} (DFTdDsC), and
the Tkatchenko-Scheffler method \cite{Tkatchenko-Scheffler} (TS)
method, and its self-consistent screened version \cite{MBD} (SCSTS).
The many-body dispersion method (MBD@rsSCS) \cite{MBD}, based on SCSTS but with explicit
many-body Dobson-B contributions collapses in calculations of materials with large
polarizabilities, including transition metals in the fourth and fifth
rows of the periodic table. We therefore instead use a recently
introduced modification of the MBD@rsSCS method, the FIA method, which
involves a more sophisticated treatment of polarizabilities by drawing
from the properties of fractional ions \cite{Gould2016-C6}.
FIA has been shown to perform as well as MBD in molecular tests, but
significantly outperforms it in strongly polarizable systems, such as
transition metal dichalcogenides, interactions involving ions, and
benzene dimers \cite{Gould2016-FI,Claudot2018-Bench,Benzene}.
The other computational methods are all based
around the two-point vdW density-functional approach of
Dion \emph{et al.} \cite{Dion}. This vdW correction is applied with the revPBE density
functional \cite{klimevs2009chemical}, the optPBE density
functional \cite{klimevs2009chemical} (optPBEvdW) and the optB88 density
functional \cite{klimevs2009chemical} (optB88vdW).  Also, in a form
modified by Lee \textit{et al.}, it is combined \cite{vdWDF2} with the
BP86 density functional \cite{PW86} (vdWDF2).

All calculations are performed using VASP 5.4.1 \cite{KressePRB1996},
where the valence electrons are separated from the core by use of
projector-augmented wave pseudopotentials (PAW) \cite{BlochlPRB1994}.
The energy cut-off for the plane-wave basis functions was set at
500~eV.  The energy tolerance for the electronic structure determinations
was set at $10^{-7}$ eV to ensure accuracy. The diversity of the
structures investigated here ensure that our results are not dependent
upon the choice of systems. We use \textbf{k}-space grids of
$9\times 9\times 3$ for graphene and boron nitride, $11\times
11\times 3$ for PdTe$_2$, and $15\times
15\times 3$ for the rest, based on energy convergence. Geometry
optimizations were performed for all structures, terminating when the
forces on all atoms fell below
0.01 eV/{\AA}. MBD and FI are calculated using the reciprocal space
implementation \cite{Bucko2016-MBD}.
The in-plane lattice parameters $a$ are kept fixed at
the respective experimental values, in accordance with previous
  work \cite{Bjorkman2012}. Small differences between the experimental
and the equilibrium theoretical $a$ lattice parameter do not
significantly affect the results for equilibrium $c_0$ and $E_b$.

\section{Results}
Our calculations of $c_0$ and $E_b$ naturally
divide the vdW methods considered
into two groups, as summarized in Table~\ref{tab:methods}: one
group tends to perform poorly for $c_0$ and
acceptably for $E_b$. The other
does the opposite. In order to quantify this classification, we
display in Fig.~\ref{fig:Rankings} the normalized mean average error
values of $c_0$ or $E_b$, given by 
$N_{c_0/E_b}(vdW) =
MAE_{c_0/E_b}(vdW)/[\frac{1}{N_{vdW}}\sum_{vdW}MAE_{c_0/E_b}(vdW)]$, where
$vdW$ labels the different $N_{\vdW}=12$
(11 tested here and SCAN+rVV10)
methods considered, and $MAE_{c_0/E_b}$ is the
mean average error across the 11 layered compounds. \textbf{For each
  of the 11 vdW methods tested here (plus results for SCAN+rVV10
  from the literature\cite{SCANrVV10}),
  the MAE for the prediction of $c_0/E_b$ is
  given by $MAE(vdW)=\sum_{i}^{11}\lvert X_{RPA}-X_{vdW}\lvert/11$,
  where $i$ iterates over the number of structures, $X_{RPA}$ is the
  benchmark $c_0$ or $E_b$ value, and $X_{vdW}$ is the calculated
  $c_0$ or $E_b$ value. According to this scheme, an accurate vdW
  method is one whose $N_{c_0}(vdW)$ and $N_{E_b}(vdW)$ are both
signficantly less than one. That is, such a method
  is able to closely reproduce the results obtained using RPA.}

The grouping discussed above is immediately obvious from this plot.
Results clearly fall into two groups of vdW methods: group I that
includes TS, DFTD3BJ, SCSTS, DFTD2, DFTD3, optB88vdW, and group
II that includes DFTdDsC, optPBEvdW, vdWDF2 and Dion. While each
group has tradeoffs, an important feature made clear by
this figure is that the
FIA methods sits at the intersection of the two, and thus achieves a
balance between energies and lattice constants.
Note that standard MBD theory fails completely for transition metal
dichalcogenides\cite{Gould2016-FI} and has been left out of these tests.

We can identify the four best methods from Fig. \ref{fig:Rankings}: FIA,
DFTdDsC, DFTD3 and optB88vdW, based on the observation that they have
the smallest $N_{c_0/E_b}(vdW)$ values. For these four methods, and for
the 11 vdW materials considered, the inset displays a scatter plot of
the values of $\Delta c_0/c_0$ and $\Delta E_b$. We identify a region
(in pink) that marks the optimal accuracy targets for each of the four
methods across the set of structures. The values $\pm 2\%$ \cite{SCANrVV10} and $\pm
5$~meV/fu (functional unit) for $\Delta c_0/c_0$ and $\Delta E_b$, respectively, are
chosen based on the following criteria: for the $c$ lattice parameter,
it is on the scale of zero-point energy effects; for $\Delta E_b$, it
is about what one expects for a typical ``registry'' difference
(e.g., between AA' and AB in MoS$_2$ -- discussed in more detail
below). The inset in the figure shows that, while no method achieves
the desired performance across even most materials, the FIA method is
the one that has the most results within the shaded circle.

So far we have focused on properties of optimal homostructures,
i.e. layered materials in which the layers have been arranged in their
lowest energy configuration. In future, the primary goal of studying
layered materials is likely to shift to \emph{heterostructures}, in
which perfect registry is impossible to achieve due to different
lattice parameters. Thus, it is important to ask whether or not
  methods are sufficiently accurate for heterostructures.
Reliable heterostructure benchmarks from
  RPA or other high-level theories are beyond current computational
limits, however, making benchmarking impossible.
Graphene/BN being a notable exception.\cite{Leconte2017-Moire}

To overcome this limitation, we test instead an important and related
property of homostructures, namely the difference in energies between
structures arranged in non-optimal stackings [illustrated in
  Fig~\ref{fig:Stacking}(a)], and the energy of the lowest energy
state. Since heterostructures involve many atoms outside optimal
registry, the ability of a method to reproduce these energy
differences will be important for accurate calculations.
Furthermore, these energies involve an interplay between
dispersive and electrostatic forces\cite{constantinescu2013stacking},
and thus methods cannot rely on any convenient cancellation of errors
near the optimal lattice point and must reproduce both with
  sufficient accuracy. Thus, although imperfect, these tests
are likely to be the best available proxy for heterostructure
physics.

We examine the influence of stacking on the contribution of each of
the 11 methods, and display the results in Fig. \ref{fig:Stacking}(b). We focus on the
following stacking configurations for graphene, hBN and MoS$_2$:
\textit{AA} in graphene, hBN and MoS$_2$, \textit{AB} in hBN and MoS$_2$
and \textit{A'B} in MoS$_2$. These cases all have reliable
  RPA benchmark data\cite{He2014,Leconte2017-Moire}, and include
both small and large energy differences to cover different physical
regimes that may be encountered in real heterostructures.
Our results show that two of the four methods already identified as good,
namely optB88vdW, and FIA, are also the best at
capturing energy differences (note, SCSTS does best on the
energy difference tests, but is worst for binding energies).
The consistent behavior of the four methods emphasizes their
general accuracy in various situations.

Finally, in our assessment of the various dispersion approaches, we
devote some special attention to PdTe$_2$, which exhibits the
interesting property that the covalent and vdW dispersion
forces \textit{compete}; according to our calculations, the
application of PBE (without any vdW dispersion model) yields a $c_0$
of 5.327 \AA, and a $E_b$ of 18.2 meV,
which are much closer to the experimental values than the
corresponding quantities for other TMC compounds when calculated using
PBE. The performance of the vdW dispersion methods in the case of this
compound is important in identifying the behavior of the dispersion
forces in an extreme case. Out of the 11 vdW methods, the ones that
have lowest errors in $E_b$ are DFTdDsC, SCSTS, TS and optPBEvdW,
while  the ones with the lowest errors in $c_0$ are DFTD2, DFTD3,
DFTD3BJ, DFTdDsC, FIA, SCSTS, TS and optB88vdW. This means that, while
the FIA method, on average achieves the tradeoff, other methods
can be more accurate in selected cases. 

Having established that the FIA method 
has the most competitive agreement with RPA
among the GGA-based methods investigated in the present Letter, we
note that recently published results on the dispersion-corrected
meta-GGA SCAN+rVV10 \cite{SCANrVV10} show it gives a superior
performance even to FIA, with about 40\% average improvements to
lattice constants and energies (no results are available for
stacking energies), despite poor performance
\cite{Bjorkman2012-Review} for PBE+rVV10 in the same
systems. However, a critical issue with SCAN+rVV10 is the
computational performance: given the complexity of the evaluation of
the kinetic energy density in meta-GGA based methods, would
SCAN+rVV10 suffer from higher computational complexity? To quantify
this, we have performed a full-relaxation calculation on a hybrid
bilayer system composed of graphene and WS$2$. This system has $3
\times 3$ WS2 and $4 \times 4$ graphene, and both calculations
started with the same initial atomic structure and with an energy cut-off of 600 eV (which is higher than the value used by rest of the calculations here, because this is required for the convergence of the meta-GGA functional). The full relaxation
of this system using SCAN+rVV10 on 64 cores required $\sim 4.1$
times the time FIA requires to perform the same calculation.

Therefore, while SCAN+rVV10 and FIA have competitive accuracies,
especially when compared to other methods tested here,
an FIA calculation takes far less time than SCAN+rVV10.
\textbf{We point out that the recent re-parametrization of the
  PBE+rVV10  method, known as PBE+rVV10L \cite{Rehabilitation} has
  been reported to yield reasonably accurate results that are
  comparable to SCAN+rVV10. But this method requires the tuning of a
  fitting parameter for different systems which makes it of limited
  applicability in general since one cannot us the same method to
  treat, e.g., a molecule adsorbed to a layered surface. A similar
  strategy was previously employed by Bj\"orkman~\emph{et al}%
  \cite{Bjorkman2012} who simply scaled VV10 energies by
  66\% to better match RPA results for layered materials.}


\section{Conclusions}
We have investigated the accuracy of 11 vdW dispersion methods for the
prediction of the geometric and energetic properties of 11
representative vdW materials, and we have found that there is a
tradeoff between the accuracy in determining the geometric and
energetic properties. Out of the 11 methods, we report that the
recently introduced FIA methods achieves the tradeoff, and that the
FIA, DFTD3, optB88vdW and DFTdDsC methods achieve high accuracy with
respect to the other methods. Two methods out of these four,
namely FIA and optB88vdW, deliver more accurate predictions compared
to the other two for 2D materials with nonequilibrium stacking
orders.

We believe that the ability of most methods to get good lattice
  parameter or energies, but not both, points to underlying problems in
  the abiliy of their polarizability models to adjust to different
  geometries (Dobson-A and -B non-additivity), especially when the
  layers are brought close to contact. Consequently, the
  damping function which connects the dispersion correction to the
  underlying exchange-correlation functional is unable to meet the
  competing demands of getting both energies and lattice parameters
  right. Only methods with very good underlying polarizability physics,
  such as FIA or SCAN+rVV10\footnote{Note, the SCAN functional
    contains good short-range dispersion contributions, but neglects
    long-range terms. It can therefore make up for deficiencies in
    rVV10, which are heavily damped near contact by the large damping
    parameter $b=15.7$.}, give sufficiently good dispersion energies near
  contact to reproduce geometries and energies together. This argument
  is supported by out-of-equilibrium results for the benzene
  dimer\cite{Benzene}.

It is interesting to note that each of the four best methods found
here represents the latest generation of a different class of vdW
method, highlighting the steady improvements in each class. Our work
thus suggests two important elements for the success of future vdW
methods: (a) achievement of the tradeoff between geometry and energy
characteristics, and (b) inclusion of the physical principles that
drive the current methods, perhaps by borrowing ``best practice'' from
methods of a different class.

Importantly, our work highlights the need to test and develop
  methods using a wide range of systems. Most dispersion methods are
  optimized and initially tested on small molecular systems, due in
  part to the availablility of high-quality benchmark data. However,
  as we have shown here this does not necessarily mean they work well
  in layered systems. Nor, presumably, when molecules are physi- or
  cheimsorpbed onto surfaces. We feel this motivates the need for
  better benchmark data of difficult systems.

Let us finally draw our attention to the most promising route
for improving dispersion force modelling.
Here we focused on generalized gradient approximation (GGA) based
approaches, due to their wide availability. The meta-GGA SCAN+rVV10
offers superior performance to any of the GGA-based approaches,
however, despite known problems with PBE+VV10 for
layered systems and a complete absence of Dobson-B
contributions. This suggests that (modified) meta-GGAs may
offer a superior starting point for dispersion corrections. Combining
the most reliable dispersion corrections here (e.g. FIA) with
meta-GGAs may thus offer the possibility of even better performance
going into the future. Progress along these lines is being pursued

\acknowledgements

This research was funded by the Australian Government through the
Australian Research Council (ARC DP160101301). Theoretical
calculations were undertaken with resources provided by the National
Computational Infrastructure (NCI) supported by the Australian
Government and by the Pawsey Supercomputing Centre funded by the
Australian Government and the Government of Western Australia.


\bibliography{Resubmit}

\end{document}